\documentclass[12pt,titlepage]{article}  

\usepackage{graphicx}
\usepackage{amssymb,amsmath}
\usepackage[numbers]{natbib}
\usepackage[titletoc,title]{appendix}
\usepackage{booktabs}
\usepackage{multirow}
\newcommand{\tabitem}{~~\llap{\textbullet}~~}
\usepackage[table]{xcolor}
\usepackage[normalem]{ulem}
\usepackage{sectsty}
\subsubsectionfont{\normalfont\large\underline}
\usepackage{caption}
\begin{document}

\begin{titlepage}
	\centering
    {\Huge \textbf{Identification of Intended Arm Movement 
    		Using Electrocorticographic Signals} \\}
	\vspace{0.5cm}
    	\vspace{2cm}
    {\large Amin Behdad $^1$, Amro Nour $^1$, Arash Zereshkian $^1$, 
    Cesar M\'arquez Chin $^{1,2}$, Prof. Milos Popovic $^{1,2,3}$\\}
	\vspace{1.5cm}

	{\normalsize \textit{$^1$The Edward S. Rogers Sr. Department of Electrical and Computer Engineering, University of Toronto} \\ 
	$^2$\textit{Rehabilitation Engineering Laboratory, Institute of Biomaterials and Biomedical Engineering, University of Toronto}\\
	$^3$\textit{Toronto Rehabilitation Institute}\\}
	\vspace{1cm}
	\vfill
\end{titlepage}

\tableofcontents

\cleardoublepage

\section{Project Description}

\subsection{Background and Motivation}

A Brain Computer Interface (BCI) is a communication system that receives neurological signals from the brain and translates them into control commands for electrical (e.g., computer mouse) and electromechanical (e.g., Wheel\-chair) devices. The development of such systems was intended originally to aid individuals with a condition called locked-in syndrome \cite{wolpaw2000brain} \cite{Sarraf_2016}. Individuals with this condition have lost all their voluntary muscle control but remain cognitively intact (i.e., mentally aware of their surroundings- can feel emotions, recognize objects/people but are unable to move). This means that they are trapped in their own bodies. The use of BCI may one day improve the independence and quality of life of people with this disability.

Recently, there has been an increased interest in exploring the use of BCIs to assist individuals with other forms of paralysis \cite{chin2007identification} \cite{sarraf2014brain}. With this shift in interest, the population that can benefit from BCIs includes individuals with high level spinal cord injuries and Cerebrovascular accidents (stroke). As a result of the added interest, the potential applications and demand for the technology have also grown. 

\begin{figure}[h]
\centering
\includegraphics[width=0.65\textwidth]{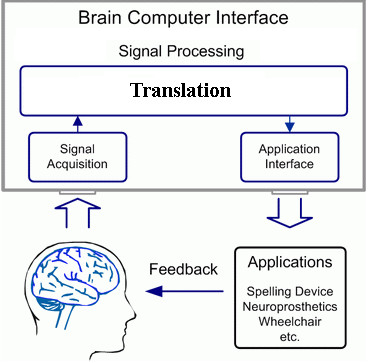}
\caption{Structural components of a BCI system \cite{pfurtscheller2006eeg}.}
\label{fig101}
\end{figure}

A general layout of a typical BCI is shown in Figure \ref{fig101}. Different mental states are detected as changes in brain activity using various recording techniques such as electroencephalography (EEG), electrocorticography (ECoG). (These recording techniques are further explained in Appendix E along with the benefits and disadvantages associated with each of them). The signals from the brain are digitized using a data acquisition system and then processed to interpret/translate them into control commands to an application interface. The interface can use the control commands to operate assistive devices such as a computer mouse, spelling device, etc. This cycle is repeated as long as there are inputs to the system. 

To have a direct brain to computer communication, patterns of brain activity are generated and then distinguished by the computer \cite{graimann2003detection} \cite{sarraf2016deep}. Two major patterns that are present in EEG and ECoG recordings are Event-Related Potentials (ERP) and Event-Related (De)synchronization (ERD/ERS). These can be observed when a person performs any voluntary movement. ERD/ERS are very important for the development of BCI systems as they can be detected before the onset of an actual movement. They also occur when a person imagines moving. This makes it possible to predict the movement without its actual occurrence, fulfilling the BCI purpose of assisting individuals who cannot perform voluntary movements \cite{sarraf2016deepad} \cite{saverino2016associative}.

This project is an attempt to create a BCI that utilizes both ERP and ERD/ERS patterns to identify different movements performed by an individual. This is part of a long term project in the Rehabilitation Engineering Laboratory of the Toronto Rehabilitation Institute and the Institute of Biomaterials and Biomedical Engineering of the University of Toronto to study the use of BCI technology to assist individuals with a spinal cord injury.

\subsection{Project Goal}

The goal of this project is to create a BCI that can identify and discriminate the ``intension'' of three different arm movements by analyzing ECoG recordings. The functionality of the BCI will be demonstrated by controlling a remote controlled car.

\subsection{Project Requirements}

Our BCI will require the design of the following two Stages:

\subsubsection{Stage 1}

The goal of this Stage is to identify, from the ECoG recordings, the ``intention'' of an individual to perform specific arm movements. 

\begin{table}[h]
\centering
  \resizebox{0.9\textwidth}{!}{
\begin{tabular}{|p{5cm}|p{8cm}|}
\hline
\rowcolor{gray} 
Functional Requirement(s) & Description  \\
\hline
\multirow{2}{12em}{Output: Predefined digital Code} & 
\tabitem Three different predefined digital codes shall be given to each successful identification of intended arm movement\\
 & \tabitem One predefined digital code shall be given to all the other arm movements\\
\hline
\end{tabular}
}
\end{table}
\begin{table}[h]
\centering
  \resizebox{0.9\textwidth}{!}{
\begin{tabular}{|p{5cm}|p{8cm}|}
\hline
\rowcolor{gray} 
Constraint(s) & Description  \\
\hline
Software: MatLab & 
Signal processing techniques used shall be coded with this software \\
\hline
Arm movement: Restricted to three & Only three movements out of all the possible arm movements shall be identified. The movements are illustrated and explained in Figure \ref{fig111} of Appendix D \\
\hline
ECoG data-files & The recognition of intended arm movements shall be implemented by keeping in mind a limited  number of trials available to us \\
\hline
Recognition failure rate: between 15\% to 20\% & This is the failure rate that our detection algorithm should abide by \\
\hline
\end{tabular}
}
\end{table}
\begin{table}[h]
\centering
  \resizebox{0.9\textwidth}{!}{
\begin{tabular}{|p{5cm}|p{8cm}|}
\hline
\rowcolor{gray} 
Objective(s) & Description  \\
\hline
Maximize detection speed & The higher the computational speed of recognizing arm movements the better \\
\hline
\end{tabular}
}
\end{table}

\subsubsection{Stage 2}

The goal of this Stage is to operate a remote control car using brain signals corresponding to different arm movements identified in Stage 1.

\begin{table}[h]
\centering
  \resizebox{0.9\textwidth}{!}{
\begin{tabular}{|p{5cm}|p{8cm}|}
\hline
\rowcolor{gray} 
Functional Requirement(s) & Description  \\
\hline
\multirow{2}{12em}{Output: Movement of car} & 
The remote control car shall move in different directions via command signals obtained from the previous Stage \\
\hline
\end{tabular}
}
\end{table}
\begin{table}[h]
\centering
  \resizebox{0.9\textwidth}{!}{
\begin{tabular}{|p{5cm}|p{8cm}|}
\hline
\rowcolor{gray} 
Constraint(s) & Description  \\
\hline
Software: LabView & 
Control of the remote control car shall be done with this software \\
\hline
Input: Single switch & The remote control car shall only operate via activating a single switch \\
\hline
Input command & The three identified arm movements shall correspond to the same command to the remote \\
\hline
Co-ordinates: Multi-directional & The remote control car must be able to maneuver in multiple directions (i.e. NNE, NE, E, W, etc.) \\
\hline
Direction changes & Direction changes are represented by state changes, which are done by switch activations. This is done continuously until the desired heading is achieved \\
\hline
\end{tabular}
}
\end{table}
\begin{table}[h]
\centering
  \resizebox{0.9\textwidth}{!}{
\begin{tabular}{|p{5cm}|p{8cm}|}
\hline
\rowcolor{gray} 
Objective(s) & Description  \\
\hline
Minimize data & The lower the amount of data used to operate the car the better \\
\hline
\end{tabular}
}
\end{table}

\subsection{Validation and Acceptance Tests}

The following will be done in order to verify that the results of our project are correct and meet the requirements:

\subsubsection{Stage 1}

\begin{figure}[h]
\centering
\includegraphics[width=0.95\textwidth]{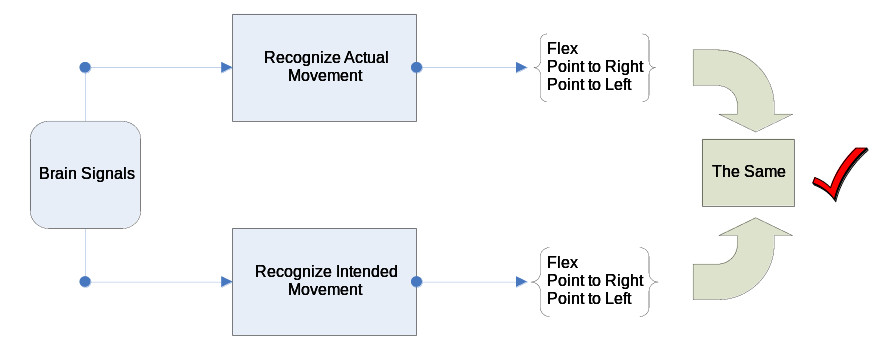}
\caption{Verification of Stage 1.}
\label{fig102}
\end{figure}

\begin{itemize}
\item Half of the available ECoG recordings will be used to implement the recognition system and the second half to test Stage 1 of our project.  Bootstrap techniques \cite{rémond1972handbook} \cite{sarraf2016robust} will be used to generate more samples of data because of low number of ECoG recordings.
\item Two instances of our recognition system will run simultaneously. Both instances will be fed with the same ECoG recordings. One instance will analyze brain signals containing activity during the execution and termination of the arm movement. The second identification system will analyze brain signals only containing activity prior to the onset of movement. The same output of each instance would mean that the system can successfully detect the intension of an arm movement (Figure \ref{fig102}).
\item In order to test the systems robustness various invalid inputs will be fed to the system to see how it would respond to ``movements'' for which it is not trained to handle.
\end{itemize}

\subsubsection{Stage 2}

\begin{itemize}
\item In order to verify the success of this Stage we will randomly feed ECoG signals to our recognition system in Stage 1 and test to see whether each input triggers a state change (i.e. direction change) in the movement of the car.
\item No other tests/verifications can be performed on this Stage due to the system not being run in real-time (we do not have access to participants). More verifiable tests could have been done if we had access to altered brain signals due to human feedback to the output of the system, rather than generated ones.
\end{itemize}

\section{Technical Design}

\subsection{Possible Solutions and Design Alternatives}

Signals from the brain entering as inputs to the BCI system need to be processed and analyzed in order to obtain useful information about them. This is done by first extracting and defining certain features for each signal and then trying to classify and make use of those features. The methods that were explored to do this are explained in the following sections:

\subsubsection{Feature Extraction Methods}

In current study of ECoG recordings of the brain, two types of brain activity patterns are mainly used to analyze the recorded data \cite{graimann2003detectionn}. The following is a brief explanation of each with some of their advantages and disadvantages:

\begin{enumerate}
\item \textbf{Event-Related Potential (ERP):} is a waveform that is generated due to changes in the amplitude of the voltage in the electrodes that are placed on the surface of the brain (i.e. a graph of Voltage vs. Time). The changes in voltage amplitude are caused by internal and external stimulations, such as sound, touch and light. In other words ERPs can be understood as reactivity patterns of a stationary system to a stimulus. This waveform is a time series that changes over time depending on the type of movement an individual performs during the recording \cite{handy2005event}.
\begin{itemize}
\item \textit{Advantage}: Considerable research is being done on this topic and the patterns generated during the movements are well known.
\item \textit{Disadvantage}: Not many of the researches focus on the waveforms generated ``before'' the actual movement, and patterns are not very well known for that portion of recording.
\end{itemize}
\item \textbf{Event-Related (De)synchronization (ERS/ERD):} are increase (ERS) or decrease (ERD) in the voltage amplitude of the signals obtained from electrodes with respect to the average voltage of many trials \cite{rémond1972handbook}. These brain activity phenomena are represented as waveforms in the form of Percent Change in Voltage vs. Time graphs, and are created for different frequencies.
\begin{itemize}
\item \textit{Advantage}: It has been proven that these patterns are capable of identifying imagined movements of an individual’s body \cite{graimann2003detection}.
\item \textit{Disadvantage}: Since the waveforms are generated by utilizing the average of many trials, many variations may be seen between each individual trial, which reduces the reliability of the system.
\end{itemize}
\end{enumerate}

\subsubsection{Classification Methods}

After the features of the brain signals are extracted by any of the methods described above, we will implement a way to detect the specific movement performed at the time of the recording. To do this the features should be compared to an ongoing brain signal, and the instances at which the specific movements occur need to be noted. This rather mathematical problem is called the ``classification process''.

A problem that every classifier needs to deal with is, given N training samples with known class labels (i.e., movements) how can one predict the class label of an unknown sample \cite{gao2007center}?

Many different classification methods are available. The following is a list of classification approaches that we consider to be more useful for our purpose.

\begin{figure}[h]
\centering
\includegraphics[width=0.75\textwidth]{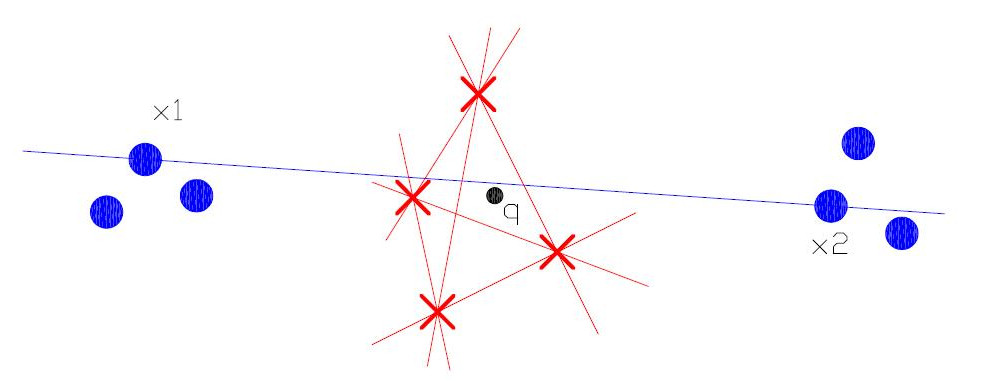}
\caption{Interpolation problem of NFL Classifier \cite{du2005pattern}.}
\label{fig103}
\end{figure}

\begin{enumerate}
\item \textbf{Nearest Neighbor Classifier \cite{zhang2006biometric}:}\\
This is a simple approach to the classifier’s problem. It consists of using the mathematical formula for the Euclidean distance between two vectors, which is $D^2 = X^2 +Y^2$. If this distance is smaller than a certain threshold then one could conclude that the unknown sample point belongs to the class of the known sample.
\begin{itemize}
\item \textit{Advantages}: Simple to implement.
\item \textit{Disadvantages}: It has a relatively high probability of error compared to other classifiers present \cite{gao2007center}.
\end{itemize}
\item \textbf{Nearest Feature Line Classifier \cite{du2005pattern}:}\\
This is an extension approach to the Nearest Neighbor Classifier. In this method instead of comparing the unknown sample point to one known sample, the comparison is done with a line. This line is obtained by interpolation of several sample points that belong to a certain class. Then the unknown sample point is said to belong to that class if its distance to the interpolated line is smaller than a certain threshold.
\begin{itemize}
\item \textit{Advantage}: The probability of the error is reduced compared to that of nearest neighbor classifier in most cases. This approach is still relatively easy to implement.
\item \textit{Disadvantage}: In some cases the interpolated line may not be a good representation of the sample points in that class. Figure \ref{fig103} shows an example of this problem. As can be seen from the figure, point ``q'' should be in the class of ``CROSSES'' however since it is closer to the interpolated line of ``DOTS'' class, the NFL classifier will consider it as a member of DOTS class which obviously is not correct.
\end{itemize}
\end{enumerate}

\subsection{System-level overview}

\begin{figure}[h]
\centering
\includegraphics[width=0.95\textwidth]{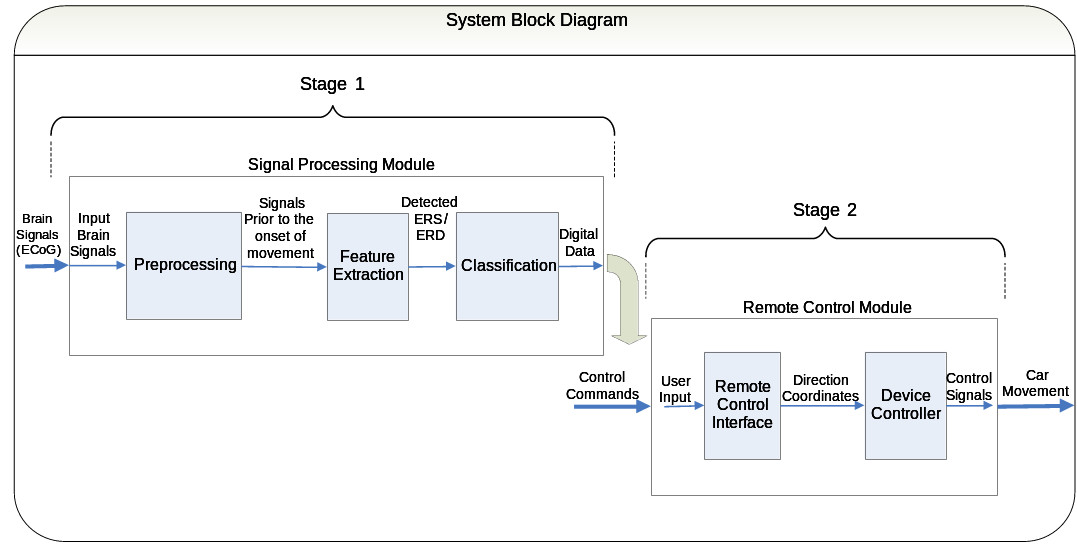}
\caption{System Level Block Diagram.}
\label{fig104}
\end{figure}

In order to implement our design we have decided to divide the project into 2 Stages as seen in Figure \ref{fig104}.

In Stage 1, the ECoG recordings are preprocessed in order to extract useful characteristics/features of brain signals, which refer to the brain activity prior to the onset of arm movements. In order to find a suitable representation of brain signals, the preprocessed signals are subsequently analyzed to detect changes following specific patterns (ERD/ERS). These patterns are then identified and classified to represent each of the three different arm movements (Figure \ref{fig111}- Appendix D). A predefined set of digital codes are then allocated to the identified arm movements.

In Stage 2, digital inputs attained from previous Stage, representing the result of the ECoG classification process, are treated as user inputs to the remote control car. The inputs fed to the remote control interface are translated into coordinates, which are used to direct the cars heading. These coordinates are then fed to the Remote Control Hardware, which outputs specific commands to the car, causing it to move in various directions.

\subsection{Module-level description}

This section outlines and explains each module of our System Block Diagram.

\subsubsection{Signal Processing Module}

\textbf{Level 0:}

\begin{figure}[h]
\centering
\renewcommand\thefigure{5.1}
\includegraphics[width=0.5\textwidth]{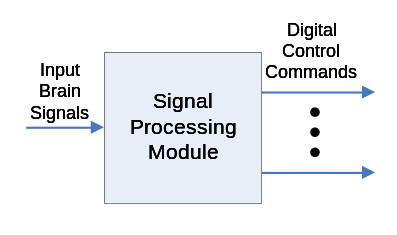}
\caption{Level 0 Signal Processing Module Functionality.}
\label{fig105}
\end{figure}

The Level 0 functionality of the Signal Processing Module is shown in Figure \ref{fig105}. The input to the system consists of brain signals and the output is a series of digital control commands. 

\begin{table}[h]
\centering
  \resizebox{0.9\textwidth}{!}{
\begin{tabular}{|p{4cm}|p{9cm}|}
\hline
\rowcolor{gray} 
\textit{Module} & Signal Processing  \\
\hline
\textit{Inputs} & - Brain signals of varying amplitudes \\
\hline
\textit{Outputs} & - Digital codes used to send out control commands to the next module \\
\hline
\textit{Functionality} & Recognition of each of the three arm movements (Figure 1- Appendix D) \\
\hline
\end{tabular}
}
\end{table}

\textbf{Level 1:}

\begin{figure}[h]
\centering
\renewcommand\thefigure{5.2}
\includegraphics[width=0.95\textwidth]{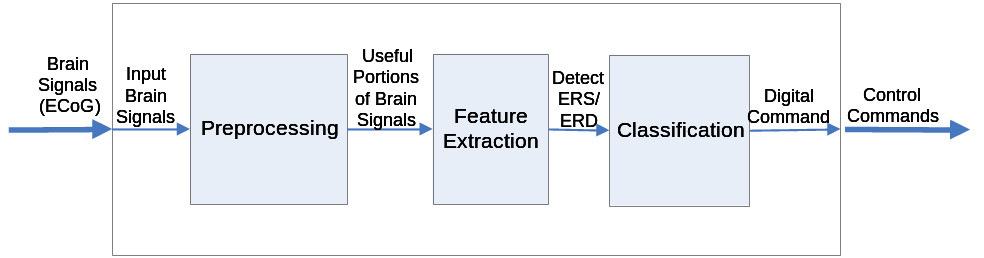}
\caption{Level 1 Signal Processing Module Functionality.}
\label{fig106}
\end{figure}

The Level 1 diagram of the Signal Processing Module is shown in Figure \ref{fig106}. It contains a Preprocessing stage that extracts useful characteristics/features from the brain signal, a Feature Extraction stage that analyses and detects variations in signal with specific patterns, and a Classification stage that identifies and classifies patterns corresponding to each of the three different arm movements.

\begin{table}[h]
\centering
  \resizebox{0.9\textwidth}{!}{
\begin{tabular}{|p{4cm}|p{9cm}|}
\hline
\rowcolor{gray} 
\textit{Module} & Preprocessing  \\
\hline
\multirow{2}{12em}{\textit{Inputs}} & - ECoG data-files \\
 & - Recorded brain signal values are in terms of voltages \\
\hline
\textit{Outputs} & - Signals prior to the onset of movement \\
\hline
\textit{Functionality} & Extracts only useful parts of the brain signal that needs to be analyzed \\
\hline
\end{tabular}
}
\end{table}

Data used in this Stage are from implanted electrodes placed over the surface of the brain (motor cortex). These data are organized in columns which represent potential differences at each instance in time.

\begin{table}[h]
\centering
  \resizebox{0.9\textwidth}{!}{
\begin{tabular}{|p{4cm}|p{9cm}|}
\hline
\rowcolor{gray} 
\textit{Module} & Feature Extraction  \\
\hline
\textit{Inputs} & - Signals prior to the onset of movement \\
\hline
\textit{Outputs} & - Identified signal patterns \\
\hline
\textit{Functionality} & Analyze and detect variations in signal patterns. These variations could be in the form of ERD or ERS. \\
\hline
\end{tabular}
}
\end{table}
\begin{table}[h]
\centering
  \resizebox{0.9\textwidth}{!}{
\begin{tabular}{|p{4cm}|p{9cm}|}
\hline
\rowcolor{gray} 
\textit{Module} & Classification  \\
\hline
\textit{Inputs} & - ECoG features \\
\hline
\textit{Outputs} & - Series of digital code \\
\hline
\textit{Functionality} & Figure out from the patterns obtained which brain signal refers to what type of arm movement. \\
\hline
\end{tabular}
}
\end{table}

\subsubsection{Remote Control Module}

\textbf{Level 0:}

\begin{figure}[h]
\centering
\renewcommand\thefigure{6.1}
\includegraphics[width=0.5\textwidth]{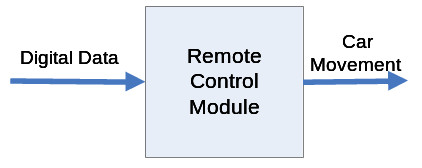}
\caption{Level 0 Remote Control Module Functionality.}
\label{fig107}
\end{figure}

The Level 0 functionality of the Remote Control Module is shown in Figure \ref{fig107}. The input to the system is a set of digital data that act as control commands or user inputs, and the output of the system is the movement of the Car.

\begin{table}[h]
\centering
  \resizebox{0.9\textwidth}{!}{
\begin{tabular}{|p{4cm}|p{9cm}|}
\hline
\rowcolor{gray} 
\textit{Module} & Remote Control  \\
\hline
\multirow{2}{12em}{\textit{Inputs}} & - Interpreted brain signals \\
 & - Stream of digital data \\
\hline
\textit{Outputs} & - Car moving in different directions \\
\hline
\textit{Functionality} & To move the car in multiple directions via interpreted brain signals corresponding to the ``intention'' of arm movement. \\
\hline
\end{tabular}
}
\end{table}

\textbf{Level 1:}

\begin{figure}[h]
\centering
\renewcommand\thefigure{6.2}
\includegraphics[width=0.95\textwidth]{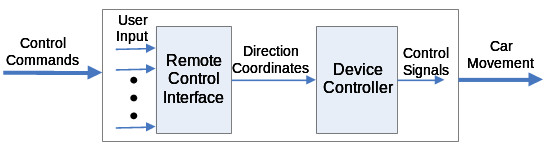}
\caption{Level 1 Remote Control Module Functionality.}
\label{fig108}
\end{figure}

The Level 1 diagram of the Remote Control Module is shown in Figure \ref{fig108}. It contains a Remote Control Interface stage that accepts user input and determines a coordinate for the direction of the Car. The Device Controller stage accepts the coordinates as input and generates control commands for the movements of car.
	
\begin{table}[h]
\centering
  \resizebox{0.9\textwidth}{!}{
\begin{tabular}{|p{4cm}|p{9cm}|}
\hline
\rowcolor{gray} 
\textit{Module} & Remote Control Interface  \\
\hline
\textit{Inputs} & - Stream of digital data \\
\hline
\textit{Outputs} & - Coordinates in multiple directions such as: NE, NNE, SE, etc \\
\hline
\textit{Functionality} & To decide from the user input which direction the car should be heading. \\
\hline
\end{tabular}
}
\end{table}
\begin{table}[h]
\centering
  \resizebox{0.9\textwidth}{!}{
\begin{tabular}{|p{4cm}|p{9cm}|}
\hline
\rowcolor{gray} 
\textit{Module} & Device Controller  \\
\hline
\textit{Inputs} & - Coordinates \\
\hline
\textit{Outputs} & - Control commands for the movement of car \\
\hline
\textit{Functionality} & To translate coordinates provided into appropriate control commands that would eventually cause the car to move. \\
\hline
\end{tabular}
}
\end{table}

\textbf{Level 2:}

\begin{figure}[h]
\centering
\renewcommand\thefigure{6.3}
\includegraphics[width=0.95\textwidth]{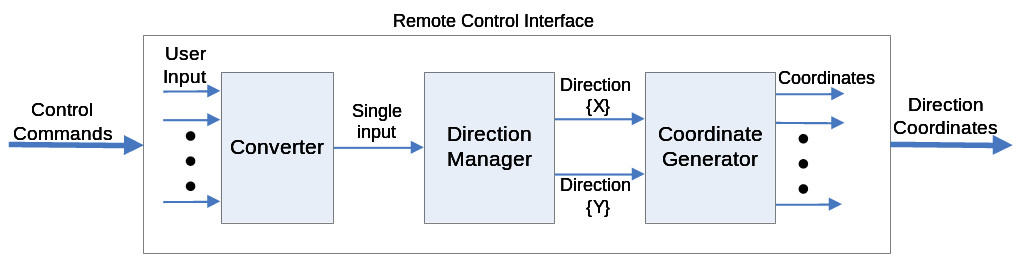}
\caption{Level 2 Remote Control Module Functionality.}
\label{fig109}
\end{figure}

The Level 2 diagram of the Remote Control Interface is shown in Figure \ref{fig109}. It contains a Converter stage that combines multiple inputs into single input, a Direction Manager that predicts and decides in which direction the car should be heading, and a Coordinate Generator that translates the desired direction into coordinates. 
	
\begin{table}[h]
\centering
  \resizebox{0.9\textwidth}{!}{
\begin{tabular}{|p{4cm}|p{9cm}|}
\hline
\rowcolor{gray} 
\textit{Module} & Converter  \\
\hline
\textit{Inputs} & - Stream of digital data \\
\hline
\textit{Outputs} & - Single 1 bit digital data \\
\hline
\textit{Functionality} & To convert multiple digital data (resembling the ``intention'' of the three arm movements as described before) into 1 bit digital data. This 1 bit data would act as a switch which has 2 states: On/Off. \\
\hline
\end{tabular}
}
\end{table}
\begin{table}[h]
\centering
  \resizebox{0.9\textwidth}{!}{
\begin{tabular}{|p{4cm}|p{9cm}|}
\hline
\rowcolor{gray} 
\textit{Module} & Direction Manager  \\
\hline
\textit{Inputs} & - Digital Code (On/Off $\rightarrow 1/0$) \\
\hline
\textit{Outputs} & - Direction of the cars heading in terms of X and Y (Cartesian Vector) \\
\hline
\textit{Functionality} & To output an appropriate heading for the car. The Direction Manager block is provided to us as a .DLL (windows environment library file) file. It is not necessary to find out how exactly it makes the decisions. \\
\hline
\end{tabular}
}
\end{table}
\begin{table}[h]
\centering
  \resizebox{0.9\textwidth}{!}{
\begin{tabular}{|p{4cm}|p{9cm}|}
\hline
\rowcolor{gray} 
\textit{Module} & Coordinate Generator  \\
\hline
\textit{Inputs} & - Directions in terms of [x, y]  \\
\hline
\textit{Outputs} & - Geographical coordinates in terms of: NNE, NE, SE, etc. \\
\hline
\textit{Functionality} & To translate the directions provided into multiple geographical coordinates. This is simply done via basic trigonometric approaches such as $\tan^{-1}(y/x)$. \\
\hline
\end{tabular}
}
\end{table}
	
\begin{figure}[h]
\centering
\renewcommand\thefigure{6.4}
\includegraphics[width=0.95\textwidth]{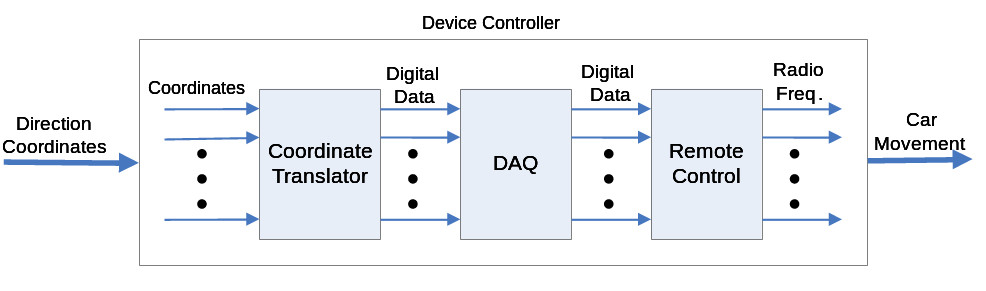}
\caption{Level 2 Remote Control Interface Functionality.}
\label{fig110}
\end{figure}

The Level 2 diagram of the Device Controller is shown in Figure \ref{fig110}. It contains a Coordinate Translator stage that converts numerical coordinates in to digital data, a DAQ (Data Acquisition) stage which serves as a software to hardware digital data transmitter, and a Remote Control stage which transmits digital data to the car.
	
\begin{table}[h]
\centering
  \resizebox{0.9\textwidth}{!}{
\begin{tabular}{|p{4cm}|p{9cm}|}
\hline
\rowcolor{gray} 
\textit{Module} & Coordinate Translator  \\
\hline
\textit{Inputs} & - Numerical data (coordinates) \\
\hline
\textit{Outputs} & - Digital data (coordinates) \\
\hline
\textit{Functionality} & To convert numerical coordinates into digital coordinates. \\
\hline
\end{tabular}
}
\end{table}
\begin{table}[h]
\centering
  \resizebox{0.9\textwidth}{!}{
\begin{tabular}{|p{4cm}|p{9cm}|}
\hline
\rowcolor{gray} 
\textit{Module} & DAQ (Data Acquisition Board NI DAQPad-6016)  \\
\hline
\textit{Inputs} & - Digital data (coordinates) \\
\hline
\textit{Outputs} & - Digital data (coordinates) \\
\hline
\textit{Functionality} & The DAQ board provides a physical link between the software and remote control hardware. With this tool, computed digital data are outputted to external hardware devices. \\
\hline
\end{tabular}
}
\end{table}
\begin{table}[h]
\centering
  \resizebox{0.9\textwidth}{!}{
\begin{tabular}{|p{4cm}|p{9cm}|}
\hline
\rowcolor{gray} 
\textit{Module} & Remote Control  \\
\hline
\textit{Inputs} & - Digital data (coordinates) \\
\hline
\textit{Outputs} & - Radio frequency \\
\hline
\textit{Functionality} & To use digital data and ultimately command the car to move in the desired heading. \\
\hline
\end{tabular}
}
\end{table}

\subsection{Assessment of Proposed Design}

This section goes over the reasons that led us choose a particular solution over others.

\subsubsection{Increased level of usability}

One of the main issues in the design of most BCIs is that their usage demands a lot of concentration from the user resulting in mental fatigue \cite{chase2006shattered}. The ultimate goal of our project is to create a brain computer interface that translates brain signals into meaningful events that could potentially increase the quality of the life of an impaired individual. Therefore, it is important for the outcome to have a significant level of usability for the users.
	
In our design we have chosen to treat all three arm movement as a trigger to change the state (i.e., direction) of the car. This basically indicates that all movements represent the same thing. This enables the user to imagine three different movements to operate the car rather than having to concentrate only on one. By reducing the concentration required by the BCI, the overall usability of the system is increased which satisfies the ultimate goal that was mentioned above. 

\subsubsection{Combining ERD/ERS and ERP patterns analysis}

Both ERD/ERS and ERP patterns have been used often in analyzing ECoG recordings of the brain activities. Many studies have shown that detection of different body movements using any of these patterns is possible \cite{huggins1999detection}. This is a good indication that ERP and ERD/ERS are the factors that should be looked at when studying ECoG recordings. 

However, using each pattern by its own introduces the following short comings and risks to the design:
\begin{itemize}
\item \textbf{ERP:} Although there has been a lot of research done in creating ERP templates that relate to specific movements \cite{huggins1999detection}, not many of them have focused on studying ERPs of imagined/intended movements. This introduces a certain level of risk to the project in a sense that the detection of imagined movement using ERP may not even be possible or may not have the level of accuracy we are looking for ($80\%-85\%$).
\item \textbf{ERD/ERS:} In contrast to ERPs, ERD/ERS patterns have been used often in detection of intended movements \cite{graimann2003detectionn}. However, ERD/ERS patterns represent averaged values over several trials. Thus individual trials may vary considerably making the use of this approach unreliable. 
\end{itemize}
From the information above, ERD/ERS and ERP patterns seem to be complimentary approaches to our design problem. Therefore an attempt to combine the two ideas will be beneficial both in terms of prediction of movement as well as increasing the reliability of the design.

\subsubsection{Nearest Neighbor (NN) vs. Nearest Feature Line (NFL) classifier}

As explained in the ``Possible Solutions'' section, NN and NFL are two possible and commonly used methods of dealing with the classification problem. Compared to NN classifier, NFL classifier is more complicated to implement but results in more accurate identification of arm movement. However, due to the time limitations and complexity of the other Stages of the project, we have decided to use the NN classifier instead. Upon successful completion of other parts of the project, we may implement the Nearest Feature Line Classifier to increase the accuracy of identifying arm movements.

\section{Work Plan}

This part of the document will focus on the management of our project. This section is composed of:
\begin{itemize}
\item Work Breakdown Structure (WBS)
\item Gantt Chart
\item Financial Plan
\item Feasibility Assessments
\end{itemize}

\subsection{Work Breakdown Structure (WBS)}

\begin{table}[h!]
\centering
  \resizebox{0.9\textwidth}{!}{
\begin{tabular}{|c|p{7cm}|c|c|c|}
\hline
\rowcolor{gray} 
\textbf{Task \#} & \textbf{Task} & \textbf{Arash} & \textbf{Amin} & \textbf{Amro}  \\
\hline
 & \textbf{Research on Brain Signal Analysis} &  &  &  \\
\hline
1 & Brain Computer Interface &  &  & R \\
\hline
2 & EEG, ECoG & R &  &  \\
\hline
3 & ERD, ERS, ERP, Classifiers &  & R &  \\
\hline
4 & Gather all Info \& Combine Research & R & R & R \\
\hline
 & \textbf{Signal Processing Techniques \& Programming Research} &  &  &  \\
\hline
1 & Spectral Signal Analysis Techniques & R & R & A \\
\hline
2 & MatLab Programming & R & R & A \\
\hline
 & \textbf{Complete and Test Stages 1} &  &  &  \\
\hline
1 & Programming for Stage 1 & R & R & A \\
\hline
2 & Testing of Stage 1 & R & R & A \\
\hline
 & \textbf{Stage 2 research} &  &  &  \\
\hline
1 & Research on DAQ Systems &  &  & R \\
\hline
2 & Assist on Stage 1 &  &  & R \\
\hline
3 & Research and finalize the Design of Stage 2 &  &  & R \\
\hline
 & \textbf{Development and Testing of Stage 2} &  &  &  \\
\hline
1 & Bringing everyone up to date & R & R & R \\
\hline
2 & Software Development of Stage 2 &  &  & R \\
\hline
3 & Hardware Development of Stage 2 & R &  &  \\
\hline
4 & Integration of Stages 1 and 2 & R & R & R \\
\hline
\multicolumn{5}{l}{\footnotesize{R = Responsible, A = Assisting}}
\end{tabular}
}
\caption{WBS}
\end{table}

\subsection{Gantt Chart}

See Appendix F: Gantt chart

\subsection{Financial Plan}

This section of the document outlines the costs of our project, which includes parts and a multi directional remote control car. The various parts are needed to modify the remote or to create a whole new interface to transmit command signals to the remote. The cost breakdown is shown below, which falls well within the limit set by the requirements.

\subsubsection{Budget Table}

\begin{table}[h!]
\centering
  \resizebox{0.9\textwidth}{!}{
\begin{tabular}{|p{4.5cm}|c|c|c|c|}
\hline
\rowcolor{gray} 
\textbf{Item} & \textbf{Cost per Unit} & \textbf{Provided} & \textbf{Quantity} & \textbf{Total cost}  \\
\hline
Remote control car & 40 & N & 1 & \$ 40  \\
\hline
Connectors & 2 & Y & 6 & \$ 0  \\
\hline
Capacitors/ Resistors/ Potentiometers & 0 & Y &  & \$ 0  \\
\hline
Oscilloscope & 0 & Y &  & \$ 0  \\
\hline
Power Supply & 0 & Y &  & \$ 0  \\
\hline
Data Acquisition Board & 0 & Y & 1 & \$ 0  \\
\hline
 &  &  & \textbf{Total} & \$ 40  \\
\hline
\end{tabular}
}
\caption{Cost}
\end{table}

\subsection{Feasibility Assessments}

\subsubsection{Skills and Resources}

Here are a series of skills and resources required to successfully complete this project:
\begin{enumerate}
\item Understanding of signal processing techniques. These include Fourier transform, power spectral analysis, etc. Currently, we are in the process of reviewing materials form various online sources.
\item Understanding of different types of brain waves and rhythms. As of now, we have a good but not complete understanding of this material.
\item Understanding of different classification techniques. We currently have limited understanding of these techniques; therefore, need to do more research on the subject.
\item Knowledge of MatLab and LabView coding. Since non of the group members have practical experience with either of the software listed, significant time needs to be dedicated to this part. We will seek help from our supervisor, the teaching assistances in the digital communication labs and online resources.
\item Familiarity with available Data Acquisition Board (National Instruments DAQPad-6016) and circuitry of remote control car. Currently we have limited knowledge on Data Acquisition Boards. There are numerous online resources and manuals available that will help us work with this module. 
\end{enumerate}

The following are the resources we have available:
\begin{enumerate}
\item Recorded ECoG brain signals from participants in the following format:\\
$ \left.
\begin{tabular}{l}
\textbullet Reaching to the right: 25 trials\\
\textbullet Reaching to the left: 23 trials\\
\textbullet Wrist Flexion: 27 trials
\end{tabular}
\right\rbrace \text{Arm Movements}$
\item Access and permission to use the available equipment in Toronto Rehab Institute Lyndhurst Centre.
\item Data Acquisition board, Oscilloscope, Power Supply and various electronic components.
\item The required software: LabView and MatLab.
\end{enumerate}

\subsubsection{Risk Assessment}

The major risk that our project could potentially face is the failure to classify/identify the ``intention'' of arm movements. The underlying reasons for this failure can be linked to:
\begin{enumerate}
\item Limited available data.
\item The utilization of the relatively new analyzing technique as described in the ``Assessments of Proposed Solution'' section of the report.
\end{enumerate}

It is impossible to attain new data from individuals at this time, but by utilizing various Bootstrap techniques \cite{rémond1972handbook} we will be able to generate more samples of data. This may increase the chances of fulfilling the project requirements.

In the event of not being able to identify the ``intended'' movements, the focus of the project can be shifted towards identification based on brain patterns ``after'' the onset of movements. This means that the design will only be relevant to individuals who have ``limited'', as opposed ``no'' muscle movements. Although this does not fulfill the project requirement, it is still a step forward for improving the quality of life of these individuals.

\clearpage

\begin{appendices}

\setcounter{figure}{0}

\section[Student-supervisor agreement form]{Student-supervisor agreement \\
form}

Our signatures below indicate that we have read and understood the following agreement, and that all parties will do their best to live up to the word as well as the spirit of it.

We agree to meet at least once every two weeks for at least half an hour to discuss progress, plans, and problems that have arisen. Before each meeting, the group will prepare a brief progress report that will form the basis for the discussions at the meeting.

If a meeting has to be cancelled by the supervisor, she/he should advise the group as early as possible. If a student cannot attend a meeting, she/he should advise members of the group as well as the supervisor as early as possible.

Both the supervisor and the students will:
\begin{itemize}
\item Inform themselves of the course expectations and grading procedure.
\end{itemize}

The supervisor will:
\begin{itemize}
\item Provide regular guidance, mentoring, and support  for his/her design project group(s),
\item Take an active role in evaluating the work and performance of the students’ by completing the supervisor’s portion of the grading forms for each course deliverable.
\item Return a photocopy of the completed grading evaluation forms to the appropriate section administrator in a timely fashion. 	
\end{itemize}

We have read and understood this agreement. Date: ................ \\
Signature of supervisor: ................................................................... \\
Signature of student: ....................................................................... \\
Signature of student: .......................................................................

\clearpage

\section{Draft B Evaluation Form (completed by ECC)}

\clearpage

\section{Report Attribution Table}

\begin{table}[ht]
\centering
\begin{tabular}{|p{7cm}|l|l|l|l|}
\hline
\multirow{2}{10em}{\textbf{Section}} & \multicolumn{4}{c|}{\textbf{Student Initials}}  \\ \cline{2-5}
 & 1. & 2. & 3. & 4. \\
\hline
 &  &  &  &  \\
\hline
 &  &  &  &  \\
\hline
 &  &  &  &  \\
\hline
 &  &  &  &  \\
\hline
 &  &  &  &  \\
\hline
 &  &  &  &  \\
\hline
 &  &  &  &  \\
\hline
 &  &  &  &  \\
\hline
 &  &  &  &  \\
\hline
\textit{All} &  &  &  &   \\
\hline
\end{tabular}
\end{table}
\noindent
\textbf{Abbreviation Codes:}
Fill in abbreviations for roles for each of the required content elements. You do not have to fill in every cell. The ``All'' row refers to the complete document and should indicate who was responsible for the final compilation and final read through of the completed document.\\
RS -- responsible for research of information \\
RD -- wrote the first draft \\
MR -- responsible for major revision\\
ET -- edited for grammar, spelling, and expression\\
OR -- other\\
``All'' row abbreviations:\\
	FP -- final read through of complete document for flow and consistency\\
	CM -- responsible for compiling the elements into the complete document\\
	OR -- other\\
If you put OR (other) in a cell please put it in as OR1, OR2, etc. Explain briefly below the role referred to:\\
OR1: enter brief description here\\
OR2: enter brief description here\\
\textbf{Signatures}
By signing below, you verify that you have read the attribution table and agree that it accurately reflects your contribution to this document.

\begin{table}[ht]
\centering
\begin{tabular}{|c|p{2cm}|c|p{2cm}|c|p{2cm}|}
\hline
\textbf{Name} &  & \textbf{Signature} &  & \textbf{Date:} & \\
\hline
\textbf{Name} &  & \textbf{Signature} &  & \textbf{Date:} & \\
\hline
\textbf{Name} &  & \textbf{Signature} &  & \textbf{Date:} & \\
\hline
\textbf{Name} &  & \textbf{Signature} &  & \textbf{Date:} & \\
\hline
\end{tabular}
\end{table}

\clearpage

\section{Figures}

\begin{figure}[h]
\centering
\includegraphics[width=0.95\textwidth]{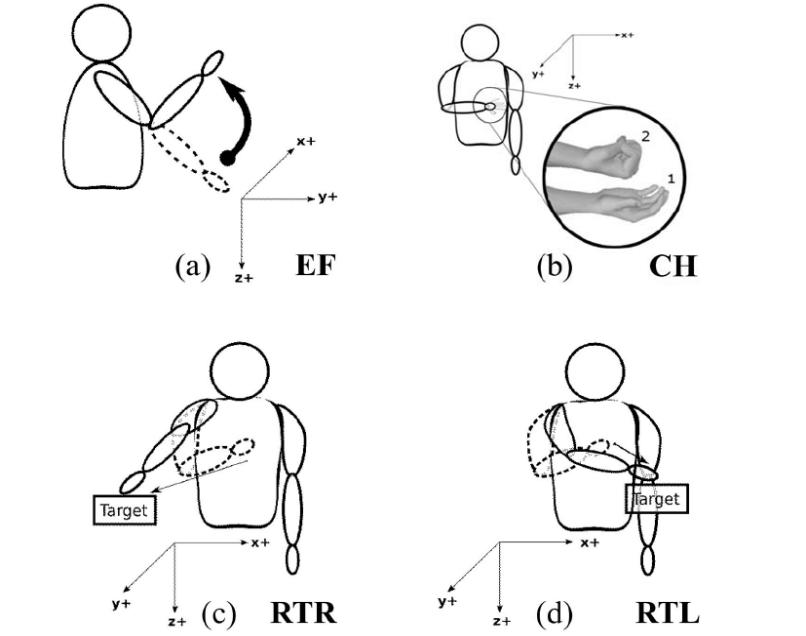}
\caption{Movements performed by the participants of the study: (a) elbow flexion (EF), (b) closing hand (CH), (c) reaching to a target placed 30 cm to the right of the individual’s midline (RTR), and (d) reaching to a target placed 30 cm to the left of the individual’s midline (RTL). [1]}
\label{fig111}
\end{figure}
\noindent
Note: 	
\par Subject 1 performed: EF, RTR, RTL
\par Subject 2 performed: CH, RTR, RTL
\clearpage

\section{Recording Techniques}

There are three major types of recoding techniques available to store electric signals that are produced by the brain. These techniques are as follows:

\textbf{Electroencephalography (EEG):} This non-invasive measurement technique is done by placing electrodes on the scalp of an individual.
\begin{itemize}
\item \textit{Advantages}: this technique is non-invasive so no surgery is required. This reduces the cost of these recordings compared to the other brain recording methods. 
\item \textit{Disadvantages}: Because the electrodes do not have a direct contact to the brain cells a lot of information is blocked which results in a poor spatial resolution on recordings.
\end{itemize}

\textbf{Electrocorticography (ECoG):}  This is an invasive technique that uses electrodes placed on the surface of the brain. The electric firing of the brain neurons are then recorded and saved for further analysis.
\begin{itemize}
\item \textit{Advantages}: Since the electrodes are placed directly on the surface of the brain, the recordings have a higher spatial resolution and higher amplitude compared to that of EEG. This enables one to measure the high frequency brain signals (i.e.gamma component). The signals at these frequencies are proved to be significant in detection of muscle movements of an individual \cite{chase2006shattered}. 
\item \textit{Disadvantages}: This method requires a surgery for electrode implantation therefore it is expensive. 
\end{itemize}

\textbf{Microelectrode Measurements:} This is an extremely invasive technique, in which micro electrodes are pierced inside the brain.
\begin{itemize}
\item \textit{Advantages}: Just like the ECoG technique, these measurements also have a high spatial resolution as well as better signal amplitude. They also measure signal from deeper portions of the brain which was not possible for either of ECoG and EEG. It also has a faster data transfer rate.
\item \textit{Disadvantages}: This method is highly invasive and very risky. Unwanted events such as probe degradation, movement and infection are probable.
\end{itemize}

\clearpage

\section{Gantt chart}

\begin{figure}[h]
\centering
\includegraphics[width=0.99\textwidth]{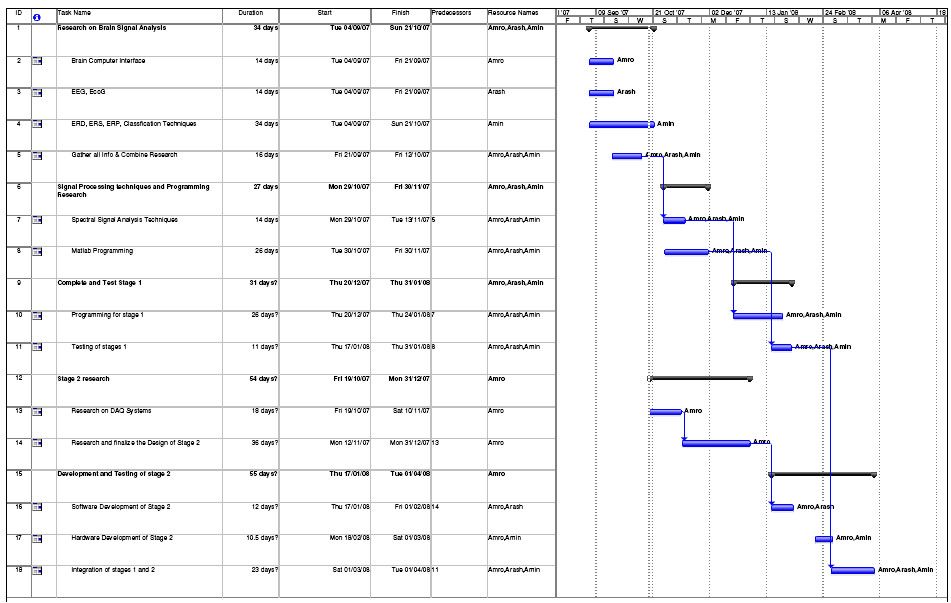}
\label{fig112}
\end{figure}

\end{appendices}

\cleardoublepage

\end{document}